\documentclass[a4paper,12pt]{nature2}
\usepackage{graphicx}
\usepackage{times}
\usepackage{color}
\usepackage{amsmath,amssymb,bm}
\usepackage{here}
\usepackage{multirow}
\usepackage{comment}
\usepackage{subfigure}
\usepackage{physics}
\usepackage{ulem}
\newcommand{\rim}[1]{\textcolor{green}{\textbf{RIM: #1}}}
\newcommand{\joe}[1]{\textcolor{blue}{\textbf{JW: #1}}}

\usepackage{xr}

\makeatletter
\newcommand*{\addFileDependency}[1]{
  \typeout{(#1)}
  \@addtofilelist{#1}
  \IfFileExists{#1}{}{\typeout{No file #1.}}
}
\makeatother

\title{Electrical observation via spin Seebeck effect of fractionalized excitations in a magnetic insulator}
\date{}

\author{
Nan~Tang $^{1}$, 
Josef Willsher $^{2}$,
Stephan Glamsch$^{3}$, 
Aisha~Aqeel$^{3}$, 
Ludwig Scheuchenpflug$^{1}$, 
Michael Schulze $^{4}$, 
Christoph Liebald $^{5}$,
Daniel Rytz $^{5}$,
Christo Guguschev$^{4}$,
Manfred Albrecht$^{3}$,
Roderich Moessner $^{2}$,
Philipp Gegenwart$^{1}$
}

\begin{document}

\maketitle

\begin{affiliations}
\item Experimental Physics VI, Center for Electronic Correlations and Magnetism, Institute of Physics, University of Augsburg, 86159 Augsburg, Germany
\item Max Planck Institute for the Physics of Complex Systems (MPI-PKS), 01187 Dresden, Germany
\item Experimental Physics IV, Institute of Physics, University of Augsburg, 86159 Augsburg, Germany
\item Leibniz Institute for Crystal Growth (IKZ), 12489 Berlin, Germany
\item EOT GmbH–Coherent, 55743 Idar-Oberstein, Germany
\end{affiliations}

\newcommand{\dto}{Dy${}_{2}$Ti${}_{2}$O${}_{7}$}
\newcommand{\one}{[1$\overline{1}$0]}
\newcommand{\gl}{${\textcolor{green}{{\maltese}}}$}
\newcommand{\VSSE}{$V_{\rm SSE}$}
\newcommand{\przro}{Pr${}_{2}$Zr${}_{2}$O${}_{7}$}

\begin{abstract}
Fractionalized excitations are among the {most striking}
signatures of emergence in quantum matter. While widely sought in frustrated magnets, their detection and characterization remain challenging, motivating the exploration of new probes. Meanwhile, \textit{Spintronics} offers versatile tools for probing spin-related phenomena. In particular, the spin Seebeck effect (SSE) converts thermally driven magnetic excitations into a voltage in an adjacent metal, providing {\it electrical} access to the underlying dynamics and transport properties. Here we employ the SSE to probe emergent magnetic monopoles in the non-collinear Ising magnet \dto{}, a rare instance of a three-dimensional fractionalized magnet. We observe an SSE signal featuring a pronounced peak at monopole proliferation, accompanied by characteristic frequency and angular dependence. Our results broaden the scope of spintronic methods for detecting exotic excitations, provide new insights into magnetic insulators generally and monopole physics specifically, and suggest the potential of quantum materials as functional interfaces. (147 words)
\end{abstract}

\newpage
\section*{Introduction}
Fractionalized excitations can emerge in strongly correlated systems, when the quantum numbers of the underlying microscopic degrees of freedom reorganize into quasiparticles carrying only a fraction of those of the original constituents. These excitations not only reveal novel forms of quantum matter, but their potentially long lifetimes and non trivial statistics also make them attractive for both fundamental studies and future device applications \cite{Giamarchi_2003, Kish_2025, Kitaev_2003}. 
They reflect strong correlations and/or an emergent gauge structure \cite{MoessnerMoore_2021}
and lie beyond a description based solely on symmetry breaking and local order parameters. Prominent examples in magnetic systems include 
spinons in quantum spin chains \cite{Tennant_1995, Giamarchi_2003, Hirobe_2017}, but they are much rarer in higher dimension. In three dimensions, emergent magnetic monopoles in spin ice\cite{CCastelnovo_2008} stand out, especially given the breadth of experimental evidence for their presence,
including neutron scattering \cite{Morris_2009, Fennell_2009}, magnetic susceptibility\cite{Snyder_2003, Matsuhira_2011, Yaraskavitch_2012, Bovo_2013, Kassner_2015, Billington_2025}, 
and flux-noise measurements\cite{Dusad_2019, Samarakoon_2022, Hsu_2024}.
Thermal conductivity measurements \cite{Kolland_2013, Scharffe_2015, Toews_2018} further suggests a role of mobile monopoles in heat transport, though the extent of monopole contribution--as opposed to non-magnetic carriers --has varied across different studies \cite{Klemke_2011, SLi_2015}.

%

Particular challenges arise when the fractionalized excitations are charge neutral~\cite{wu_2024}: charge transport -- so crucial in the discovery of fractionalization in $d>1$~\cite{tsui1982,laughlin1983}
--  is then insensitive to them. Thermal transport is still a valuable probe, but its effectiveness can in some settings be limited by phonons and other background contributions, making it difficult to isolate the magnetic signal of interest. These realities underscore the need for alternative experimental probes of fractionalized states in magnetic insulators.

In this work, we employ the spin Seebeck effect (SSE) \cite{Uchida_2008, Bauer_2012, Aqeel_2016 
} as a complementary probe of the thermally driven magnetic response, and use it to open a new window on the physics of high-dimensional fractionalization. The SSE arises from a temperature gradient across the interface between a magnetic insulator and a nonmagnetic metal with strong spin–orbit coupling, such as Pt. This temperature gradient thermally drives magnetic excitations in the insulator, resulting in angular momentum transfer across the interface and generating a pure spin current within the Pt. This spin current is then converted into a measurable charge signal via the inverse spin Hall effect (ISHE). 

As such, the SSE provides a sensitive probe of fundamental properties of a magnet, with ISHE readout enabling detection of signals down to the tens of nanovolt level. Notably, it
simultaneously probes magnetic order and excitations, selectively probing magnetic rather than, say, phononic, contributions to transport. 
To date, SSE has been observed widely from conventional magnets, 
where the response is often magnonic, to diverse quantum magnets, involving 
spinons, triplons, magnon-polarons, etc \cite{Kikkawa_2023}. 

We find qualitatively different SSE signatures in spin ice compared to either conventional ordered- or para-magnets. Spin ice is a frustrated Ising magnet on a pyrochlore lattice, characterized by the ``2-in, 2-out'' ice rule (Fig.~1{\bf a}). A single spin flip then creates two topological defects, a monopole–antimonopole pair (``3-in, 1-out'' / ``1-in, 3-out'') \cite{CCastelnovo_2008}. Successive Ising spin flips allow the monopole to hop without additional energy cost \cite{Ryzhkin_2005,Jaubert_2009, Morris_2009}.
These spin flips lack coherent spin dynamics, being dictated by the effective mobilities of spin flip (monopole) excitations \cite{Ryzhkin_2005}, rather than, say, the conventional magnons in ordered Heisenberg magnets with a well-defined quasiparticle energy. In particular, we report prominent features due to the fractionalized monopoles and the tuning of effective dimensionality that appear at low temperatures. We present a simple picture which accounts for the gross features of the dependence of the SSE signal on temperature as well as field strength and direction, and allowing us to isolate novel features due to fractionalized transport. 


\section*{Results}
As illustrated in Fig.~1{\bf b}, our device consists of a single crystal of \dto{} with a thin-film Pt Hall bar deposited on its surface. The Pt strip serves two purposes: when an AC current $I$ is applied, it locally heats the sample surface, thereby generating a temperature gradient across the sample; simultaneously, it acts as a detector, measuring the resulting SSE voltage. The temperature gradient, estimated to be approximately 0.1 K (see Methods), is applied along the \one{} direction, which aligns with two of the tetrahedral sublattices, allowing efficient spin propagation. 
Meanwhile, the magnetic field is applied in-plane along the [111] direction, known for its characteristic metamagnetic transition associated with monopole proliferation. 
For further experimental details, see the Methods section. 

\subsection{Field dependence of magnetic susceptibility and spin Seebeck voltage\\}
Since the SSE is closely linked to the magnetization, we first present the magnetic field ($B$) dependence of the magnetization ($M$) and magnetic susceptibility ($dM/dB$) at 1.4 K for $B \parallel [111]$ in Dy$_2$Ti$_2$O$_7$, as shown in Fig. 2{\bf a}. Magnetization $M$ first increases rapidly up to 0.3 T as one of four spins per tetrahedron fully polarizes along the [111] field, and then develops a shoulder associated with the kagome-ice plateau (Fig.~1{\bf a}).
With increasing magnetic fields, the system undergoes a metamagnetic transition into the fully polarized “1(3)-in, 3(1)-out” state, signaling the creation and proliferation of magnetic monopoles. 
This $B$-dependent behavior is well established in previous studies of spin ice systems both experimentally and theoretically\cite{Matsuhira_2002, TSakakibara_2003, Moessner111_2003, SVIsakov_2004}. 
The inset shows $dM/dB$ as a function of $B$ at selected temperatures. The peak associated with the metamagnetic anomaly diminishes with increasing temperature and disappears above 4 K.

Next, we present the main experimental result: $B$ dependence of the spin Seebeck voltage ($V_{\rm SSE}$) in Pt/\dto{} devices, shown in Fig.~2\textbf{b}. 
Weak SSE signals emerge around 15 K, as evidenced by the odd-in-$B$ response. This agrees with the symmetry of the SSE, in which the induced voltage $V_{\rm SSE}$ is proportional to $\nabla T \times \mathbf{M}$. Down to 7.8 K, \VSSE{} scales with the magnetization (Extended Data Fig.~1), consistent with the behavior expected for a high-temperature paramagnet \cite{SWu_2015, wang2025} and commonly observed in SSE systems more broadly. At $T= 4$ K as the system enters the spin ice regime, however, \VSSE{} no longer scales simply with $M$. Instead, a small but distinct peak develops near 0.3 T,  marking the onset of the kagome-ice state. 
Notably, the amplitude of this peak is only weakly temperature dependent. 
At $T= 2.4$~K, a second peak appears near 1.1~T, corresponding to monopole proliferation. This peak becomes more pronounced at 1.4 K, in agreement with the magnetic susceptibility $dM/dB$ (inset of Fig.~2\textbf{a}).

We confirm the thermal origin of these SSE peaks by showing that their amplitude scales with $I^2$ (Extended Data Fig.~2). Furthermore, we measured the Ta/\dto{} device under similar conditions. The key features were reproduced in Ta/\dto{}, but with a sign reversal, consistent with the opposite spin Hall angle of Ta compared to Pt, thereby confirming the spin origin of the signals. As a control, Extended Data Fig.~3\textbf{a} shows the $V_{\rm SSE}$ measured on a non-magnetic Pt/SiO$_{\rm x}$ device. No SSE signal is detected, further confirming that the observed effects originate from spin excitations in the \dto{} substrate.

We now discuss the physical origin of the field-induced peaks in \VSSE{} observed in \dto{}. 
A defining feature of our data is that these peaks dominate the low-temperature field dependence, setting \dto{} apart from previously studied SSE systems, where such features typically appear only as secondary contributions on top of a large background SSE signal and arise from
mechanisms, such as magnon-polaron formation\cite{TKikkawa_2016, Ramos_2019, Xing_2020, Junxue_2020},
or the competition between spin canting and magnon mode hardening\cite{He_2025}.
In contrast, all peaks in \dto{} can be understood within the framework of spin ice.
The peak near 0.3 T corresponds to the crossover from 3D spin-ice into the 2D kagome-ice \cite{Matsuhira_2002, TSakakibara_2003, Moessner111_2003, SVIsakov_2004}, 
while the peak near 1.1 T is associated with the metamagnetic anomaly. 
Notably, the width of the $V_{\rm SSE}$ peak is substantially narrower than that of $dM/dB$ at 1.4 K, and its amplitude follows a $1/T$ dependence (Extended Data Fig.~4\textbf{a, b}). 

\subsection{Frequency dependence of spin Seebeck voltage\\}
To further investigate the peaks observed in \VSSE, we performed frequency ($f$) sweeps at selected magnetic fields, shown in Fig.~3. Remarkably, a broad peak centered around 0.5 $\sim$ 1 kHz appears at $B = \pm 1.1$ T, corresponding to the metamagnetic anomaly.
%
%
Furthermore, the interfacial conversion between the spin dynamics in \dto{} and the charge current in Pt is expected to be effectively instantaneous within our measurement frequency range, as ultrafast studies on Y$_3$Fe$_5$O$_{12}$ (YIG) systems have demonstrated sub-picosecond to picosecond buildup of the spin Seebeck signal\cite{Kimling_2017, Seifert_2018}. In addition, a frequency sweep performed on a Pt/SiO$_x$ control device (Extended Data Fig.~3{\bf b}) show no comparable frequency dependence.
We therefore exclude the possibility that the observed kHz-scale dispersion is due to interfacial processes or instrumental artifacts. 

The observed $f$ dependence differs qualitatively from that of conventional magnetic insulators such as YIG, where the SSE typically exhibits a low-pass response: the signal remains nearly constant at low frequencies and rolls off only in the MHz range, consistent with magnon--phonon interaction times of a few hundred picoseconds in thin films\cite{Schreier_2016}. 
In \dto{}, by contrast, the $f$ dependence of SSE reveals a strongly field-dependent response: at 0.3 T, there is almost no $f$ dependence, while at 1.1T, \VSSE{} displays a resonance like-enhancement.
This broad maximum centered at drive frequencies $f = 0.5\text{--}1~\mathrm{kHz}$ reflects the system's response to thermal modulation at $2f = 1\text{--}2~\mathrm{kHz}$, since the SSE is detected at the second harmonic 
of the applied AC current. 
This corresponds to a characteristic timescale of $\tau \sim [2\pi(2f)]^{-1} \sim 0.08\text{--}0.16~\mathrm{ms}$. 
Such a peak-like $f$ dependence is not observed in other compounds to the best of our knowledge.

\subsection{Angular dependence of spin Seebeck voltage\\}
Because thermal transport
is expected to be sensitive to the magnetic structure, we further performed in-plane rotation measurements, presented in Fig. 4. All the data in Fig. 4 is shown after an 8 degree correction in measured rotator angle to match the crystallographic basis.
Our default SSE configuration presented in Figs. 2 and 3 corresponds to the nominal $90^\circ$ in Fig. 4, associated with $B \parallel [111]$.
At $B = 0~\mathrm{T}$, there is no sample magnetization and
no angle-dependent response is observed.
At $B = 0.3~\mathrm{T}$, \VSSE{} is non-zero and
follows a
sinusoidal dependence, 
as expected for the inverse spin Hall geometry, and as also observed in the spinel CoCr$_2$O$_4$ \cite{Aqeel_2025}.
Upon further increasing the field to $B = 0.6~\mathrm{T}$, the response begins to deviate strongly from a sinusoidal form with additional structures developing near $0^\circ$, $20^\circ$, and $90^\circ$.
At fields approaching the metamagnetic anomaly, $B = 0.9~\mathrm{T}$ and $1.1~\mathrm{T}$, \VSSE{} exhibits pronounced peaks at the same angles. 
These additional features reflect the complex geometry of the pyrochlore lattice. As shown in the bottom panel of Fig. 4, taking $[11\bar{2}]$ as $0^\circ$, the $[11\bar{1}]$ and $[111]$ directions correspond to $19.5^\circ$ and $90^\circ$, respectively, indicating that these peaks arise when the magnetic field aligns with these crystallographic directions.
Corresponding peak reversal is observed upon $180^\circ$ inversion on each angle.

Interestingly, although $[111]$ and $[11\bar{1}]$ are crystallographically equivalent and would therefore be expected to yield similar magnetization, one might anticipate a larger \VSSE{} response for $[111]$ because the spin Hall geometry is better satisfied near $90^\circ$. However, the peak amplitudes at these two angles are comparable, possibly due to sample misalignment.
Moreover, \VSSE{} at $0^\circ$ and $180^\circ$ exhibits the \emph{largest response} of any orientation,
despite being strongly suppressed by the spin-Hall geometry. 
This suggests a mechanism for the dramatic enhancement of the SSE signal in this geometry, discussed further in the following section.

\section*{Discussion}
We  discuss the physical origin of the distinctive $B$-, $f$-, and $\theta$-dependent features within a unified theoretical framework.
Generally,
the SSE signal measures a spin current $I_s$ injected into the surface of a magnetic insulator, which depends on several 
thermodynamic and transport processes,  including 
changes in the magnetic  state as a function of applied field and temperature, as well as the dynamics and transport of spin excitations away from the interface.

In order to identify and disentangle these effects, we employ a linear response description of the SSE \cite{oyanagi2023,wang2025}.
The injected spin current $I_s$ is driven by a temperature difference, $\Delta T_{\mathrm{int}}$, between the magnetic carriers in the insulator and the {conduction electrons in} the metal at the interface.
We assume a limit of weak coupling $J'$ between the metal and insulator and small $\Delta T_{\mathrm{int}} \ll T$, where the spin current is proportional to \cite{adachi2013}

\begin{equation}\label{eq:sse_Is_nnchichi}
    I_s \propto -J'^2 \frac{\Delta T_{\mathrm{int}}}{T^2}  \int \frac{\dd{\omega}}{2\pi}  
    \frac{\omega^2}{\sinh^2(\omega/2T)}
    \chi''_P(\omega) \, ,
\end{equation}
with $\chi''_P(\omega)$ the imaginary part of the spin susceptibility in the GHz frequency range \cite{wang2025}. 
Varying the magnetic state at the interface  causes a change in $\chi''_P(\omega)$. Additionally, changes in temperature, heating rate, sample thickness, spin-phonon coupling, or mobility of magnetic carriers  affect the interfacial temperature difference $\Delta T_{\mathrm{int}}$.
The SSE can thus probe not only thermodynamic properties but also transport-related quantities, such as thermal propagation lengths \cite{Kikkawa_2023}.

Our model of the SSE signal in spin ice has two main ingredients.
First, the interfacial spin injection in the linear response regime \cite{oyanagi2023} is 
modeled using a single-tetrahedron approximation, as described in the Supplementary Information (SI). 
Second, we define $\mu$ as the characteristic mobility of the excitations in the bulk insulating magnet. In the steady state, this mobility controls the rate at which the thermal energy injected into the spin systems is transported away from the boundary into the bulk.
For incoherent Metropolis dynamics in Ising-like systems, $\mu$ generically scales inversely with temperature, $\mu \sim 1/T$ \cite{Ryzhkin_2005, castelnovo2011}.
We assume that the interfacial temperature difference is proportional to mobility, $\Delta T_{\mathrm{int}} \sim \mu$ and is independent of the applied magnetic field $B$.
 
We begin by discussing the high temperature response $T \ge 8$ K, well above the spin ice temperature scale. In this regime, the theoretical model with no free parameters describes \VSSE{} well, as shown in Fig. S3 in the SI. 
Upon lowering the temperature,
it captures the emergence of 
the metamagnetic \VSSE{} peak at $\pm 1.1$ T, and correctly reproduces the $1/T$ scaling of this feature (Extended Data Fig.~4\textbf{b}, and Fig. S4 in the SI).
However, this model considerably underestimates the height and overestimates the width of this metamagnetic \VSSE{} peak, 
as it predicts only a hump that is not clearly distinguishable from the low-$B$ peak.
Likewise, at intermediate fields of $\pm 0.6$ T, corresponding to the kagome-plateau regime, this simple model does not reproduce the low-$T$ suppression of the signal (or the sign change at the lowest temperature).

These discrepancies immediately point towards two {\it dynamical} features not accounted for in our simple model, underpinning monopole motion and hence field-dependent magnetic transport. These are
%
the field-dependent activated nature of the monopole hopping and the varying timescales of spin flips in an external magnetic field.
To understand the first, note that in the kagome ice regime, creating a virtual monopole-antimonopole pair \cite{mostame2014} costs an energy $\Delta E \sim |B-B_c|$. 
Thus, this energy cost suppresses monopole mobility \cite{castelnovo2011} $\mu\sim \mu_0 e^{-\Delta E/k_B T}$ away from $B_c$ at low temperatures, thereby reducing the SSE signal.
By contrast, at $B=B_c$, $\Delta E$ vanishes, allowing the monopole mobility to remain large.
Because the SSE signal is sensitive to this mobility, it is enhanced only in a narrow field window around $B_c$, providing a natural mechanism for the \VSSE{} peak to be narrower than thermodynamic features, such as the broader peak in $dM/dB$.
Regarding the second, the strong enhancement of the peak at $1.1$ T can similarly reflect a field-dependent speeding up of the intrinsic spin-flip dynamics \cite{Snyder_2003, matthews2012} by enhancing the local transverse field component \cite{Tomasello_2019,Hallen_2022}. 
This mechanism is also key to interpreting the enhancement of the signal as a function of field direction, shown in angular dependence data in Fig. 4.
Orienting the field to the $[11\bar{2}]$ direction causes the spin chains along $[110]$ to experience the strongest possible transverse field. 
Such a geometry allows the characteristic timescale of magnetic transport, and hence the mobility of spin excitations, to be tuned by the field strength.
Because there is no metamagnetic transition in this orientation, this geometry would isolate the dynamic transport contribution to the SSE signal.

The driving-frequency dependence of the signal is a complementary probe of the timescales underpinning the dynamics of the magnetic excitations. 
In particular, we recall that the largest \VSSE{} response at this field strength occurs when the driving frequency approximately matches the characteristic monopole hopping rate of classical spin ice, as measured
by AC susceptibility\cite{Snyder_2004, Matsuhira_2011, Yaraskavitch_2012, Kassner_2015, Eyvazov_2018} and flux-noise measurements\cite{Dusad_2019, Hsu_2024}.
These studies report macroscopic magnetic relaxation times on the order of 1 ms at relevant temperatures, while Hallén \textit{et al.}\ \cite{Hallen_2022} identified a shorter microscopic timescale of 85 $\mu$s, from modeling anomalous magnetic-noise spectra\cite{Samarakoon_2022}, even closer to the characteristic timescale probed in our experiment.
This spin flip timescale is relevant to monopole motion, and therefore to the capacity of a monopole transport current to carry the energy injected by the interfacial spin current away from the boundary. 
Here, further work in modeling the quantitative effect of an external magnetic field on spin flip timescales and consequently on the SSE response is desirable.
%
%
%

Finally, we note that linear response theory presented here does not capture the sign inversion of the SSE signal at 1.4 K. This deviation is striking, and it also seems to be involved in giving the metamagnetic peak an asymmetrically narrowed shape. The origin of this feature remains unclear. We note that a similar phenomenon has been observed in 1D spin chains \cite{Hirobe_2017}, where it was ascribed  to interactions between spinons \cite{wang2025}, raising the question whether the feature here might likewise reflect the effect of monopole interactions on  transport in kagome ice.

In conclusion, these SSE data in \dto{} reflect a highly tunable combination of not only thermodynamic but also dynamical and transport physics of the magnetic excitations, with potential presence of cooperative effects. We therefore underscore the promise of this experimental technique in probing fractionalized classical and quantum phases of matter, even in insulating magnets.
We briefly make a comparison with thermal conductivity, another powerful probe of transport. Whereas the transport character of SSE is somewhat less direct, especially in  measurement geometries as those adopted in this study, early studies on YIG have shown that this interfacial readout technique can still carry information about bulk transport \cite{Rezende_2014, Agrawal_2014, Guo_2016}, and our experimental data likewise point to a transport-based interpretation. 
A central strength of SSE is its selectivity to magnetic transport signals, obviating the need to isolate out a potentially dominant phonon contribution to the thermal conductivity \cite{sutcliffe2022}.
From a technical point of view, frequency and angle dependence can also be incorporated more readily into an SSE setup due to their electrical detection scheme.
As noted above, in spin-ice systems in particular, this can additionally enable tuning of the effective dimensionality (i.e., from 3D spin ice to 2D kagome ice and 1D Ising chains) alongside microscopic dynamics (through the strength of the transverse field).

Our work appears to be the first detection  of three-dimensional fractionalized excitations generally, and their dynamical and transport properties specifically, via the spin Seebeck effect. Notably, this provides an {\it electrical} handle on fractionalization in an insulating magnet \cite{wu_2024}. 
Our results point at the capacity of quantum materials as functional interfaces that enable spintronic functionalities, including deconfined spin transport and information transduction through spin–charge interconversion. Looking ahead, this work lays the foundation for future devices in which fractionalized excitations—such as those in quantum spin liquids—actively mediate signal generation and transport. Harnessing these excitations in magnetothermal response introduces new design opportunities for spintronics beyond conventional magnon-based paradigms, where exotic quasiparticles mediate coupled spin, heat, and charge dynamics.

\begin{addendum}
\item We thank T. Sakakibara, A. Jesche, A. Herrnberger, M. Althammer, H. Huebl, K. Uchida, S. Wu, A. Chanda for helpful discussions. Work at Augsburg is supported by the Alexander von Humboldt Foundation and the Deutsche Forschungsgemeinschaft (DFG, German Research Foundation) through projects 492547816 (TRR360), 492421737, and 528001743, the cluster of excellence ct.qmat (EXC 2147, project-id 390858490) and  SFB 1143 (Project-ID No. 247310070). We gratefully acknowledge the funding for the single crystal growth development of \dto{} within the cooperation project “Pyrochlore crystals for compact optical isolators” provided by the Federal Ministry for Economic Affairs and Energy (former BMWi, now BMWE) due to an enactment of the German Bundestag under Grant No. ZF4728901RE9 and ZF4730501RE9 within the Central Innovation Program for SMEs (ZIM).

\item[Contributions] 
N.T. and P.G. conceived the project, N.T. designed and organized the experiments. M.S. led the single-crystal growth with support from C.L., D.R., and C.G.. N.T. oriented the samples and polished the surfaces. N.T., S.G., and L.S. performed Atomic Force Microscopy (AFM) imaging to evaluate surface roughness. S.G. carried out optical lithography and Hall bar deposition. N.T. conducted magnetization measurements, and N.T., A.A., and L.S. performed spin Seebeck measurements. N.T. analyzed the data and prepared the figures. J.W. and R.M. performed theoretical modeling and prepared  figures. N.T., J.W., and R.M. wrote the manuscript with input from all authors.

\item[Competing Interests] The authors declare that they have no competing interests.
\item[Corresponding authors]
Correspondence and requests for materials should be addressed to N. Tang. (email:nan.tang@uni-a.de).
\end{addendum}

\newpage
\begin{figure}[ht]
 \begin{center}
 \includegraphics[keepaspectratio, scale=0.63]{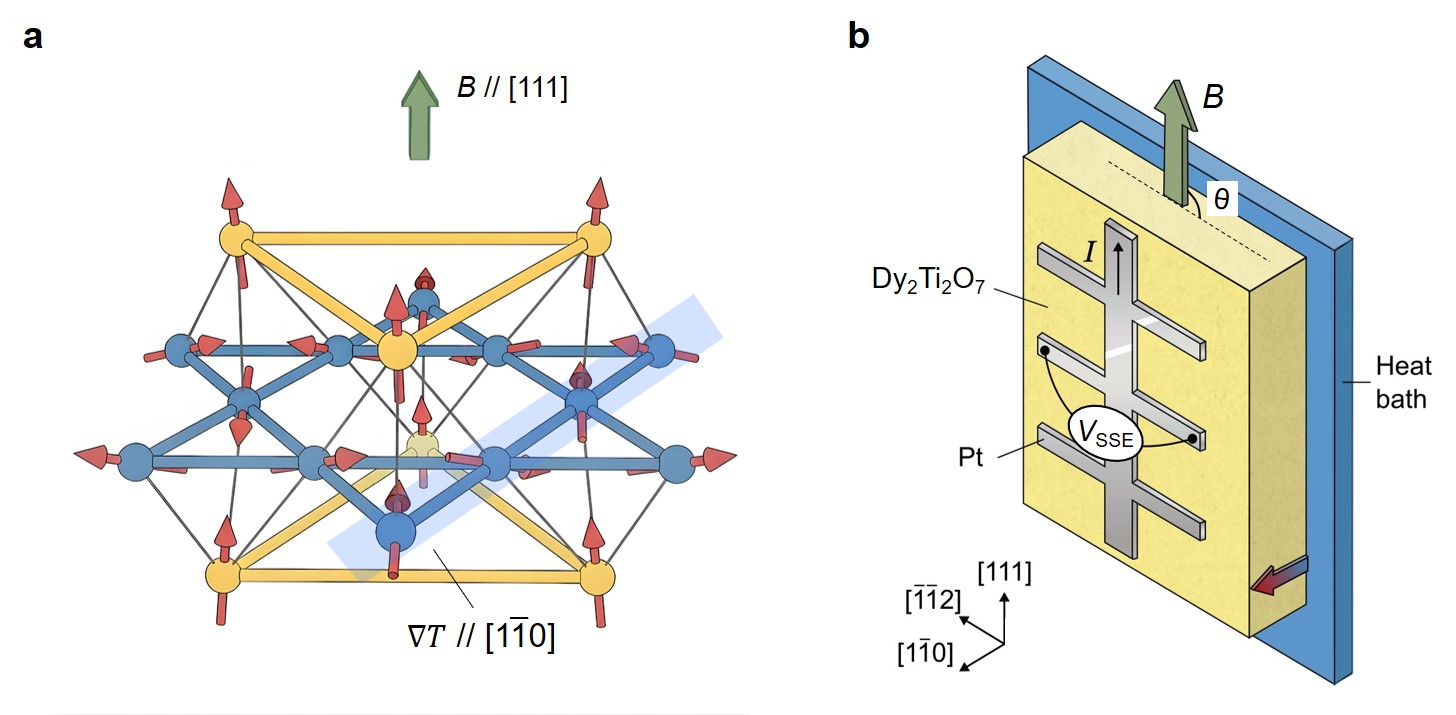}
 \end{center}
\end{figure}
 \normalsize{{\bf Figure 1 $|$ Schematic of the experimental geometry of the pyrochlore oxide \dto.}
 {\bf a}, Illustration of the pyrochlore lattice in spin ice, with crystallographic directions [111] and \one{} indicated. The magnetic moments (Ising spins) are constrained to lie along the local [111] axes of each Dy$^{3+}$ ion, forming a short-range spin-ice configuration characterized by the macroscopically degenerate ``2-in, 2-out" ice rule. The pyrochlore lattice can be viewed as an alternating stacking of triangular and kagome layers along the [111] direction. The spins on the triangular layers are easily polarized by a small magnetic field applied along [111], leaving the remaining three form a kagome-ice network satisfying the “1(2)-in, 2(1)-out” constraint. In this kagome ice state, the sixfold “2-in, 2-out” degeneracy is reduced to threefold per tetrahedron.
 {\bf b}, Schematic of the measurement configuration used for spin Seebeck effect (SSE) measurements. The device consists of a bulk \dto{} single crystal (yellow block) with a 4-nm-thick Pt Hall bar deposited on top. An AC current is applied along the Pt strip, which is oriented along the [111] direction, thereby generating a temperature gradient $\nabla T$ along the perpendicular \one{} direction. With the magnetic field applied along [111], the spin Seebeck voltage $V_{\rm SSE}$ is detected along the direction perpendicular to both the magnetic field and the thermal gradient, consistent with the orthogonality condition for SSE. We refer to this configuration as the nominal $\theta = 90^\circ$; unless otherwise specified, all measurements were performed in this configuration. 
For the angular-dependence measurements, $\theta$ was varied by rotating the sample in-plane using a rotator.
 }
\label{Fig:Fig1}

\newpage
\begin{figure}[ht]
 \begin{center}
 \includegraphics[keepaspectratio, scale=0.55]{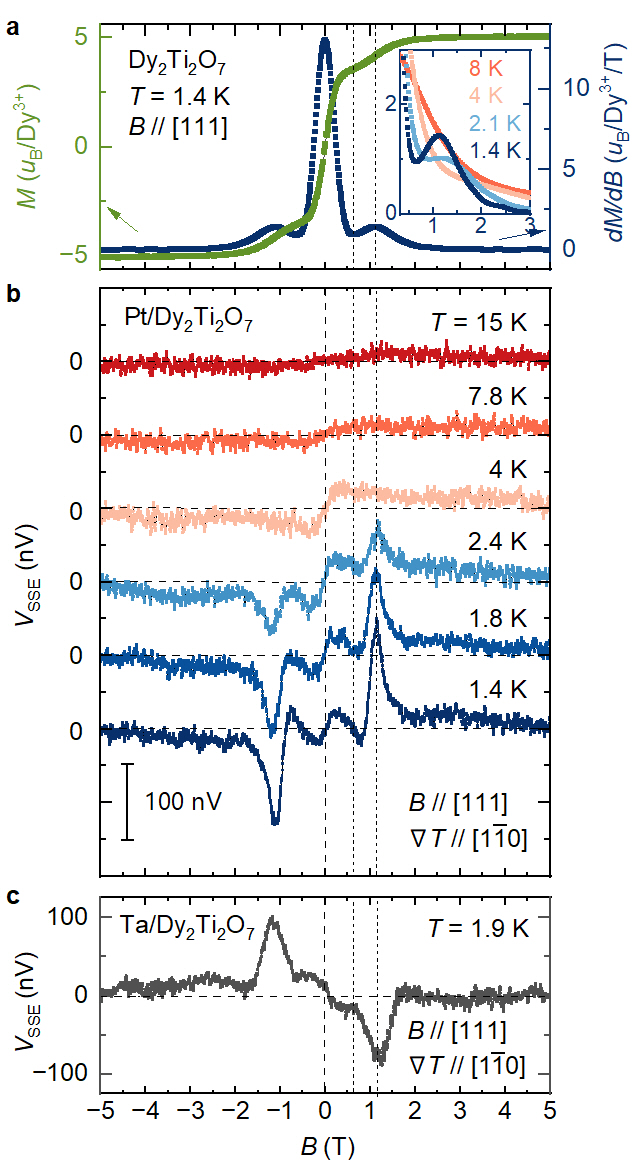}
 \end{center}
\end{figure}
 \normalsize{{\bf Figure 2 $|$ Magnetic field $B$ dependence of Magnetization $M$ and spin Seebeck voltage $V_{\rm SSE}$ in \dto.}
 {\bf a}, Magnetization $M$ (left axis) measured at 1.4 K as a function of magnetic field $B$ applied along the [111] direction. The corresponding magnetic susceptibility $dM/dB$ is shown on the right axis. Inset: $dM/dB$ curves at selected temperatures.  
 {\bf b}, $B$ dependence of the spin Seebeck voltage $V_{\rm SSE}$ of Pt/\dto{} device at selected temperatures. One major grid division corresponds to 100 nV. 
 {\bf c}, $B$ dependence of $V_{\rm SSE}$ measured on a Ta/\dto{} device under similar conditions.
 The two finely dashed lines indicate characteristic magnetic field positions at 0.6 T and 1.1 T. The coarser dashed lines mark the zero points of both the $x$- and $y$-axes.
}
\label{Fig:Fig2}

\newpage
\begin{figure}[ht]
 \begin{center}
 \includegraphics[keepaspectratio, scale=0.4]{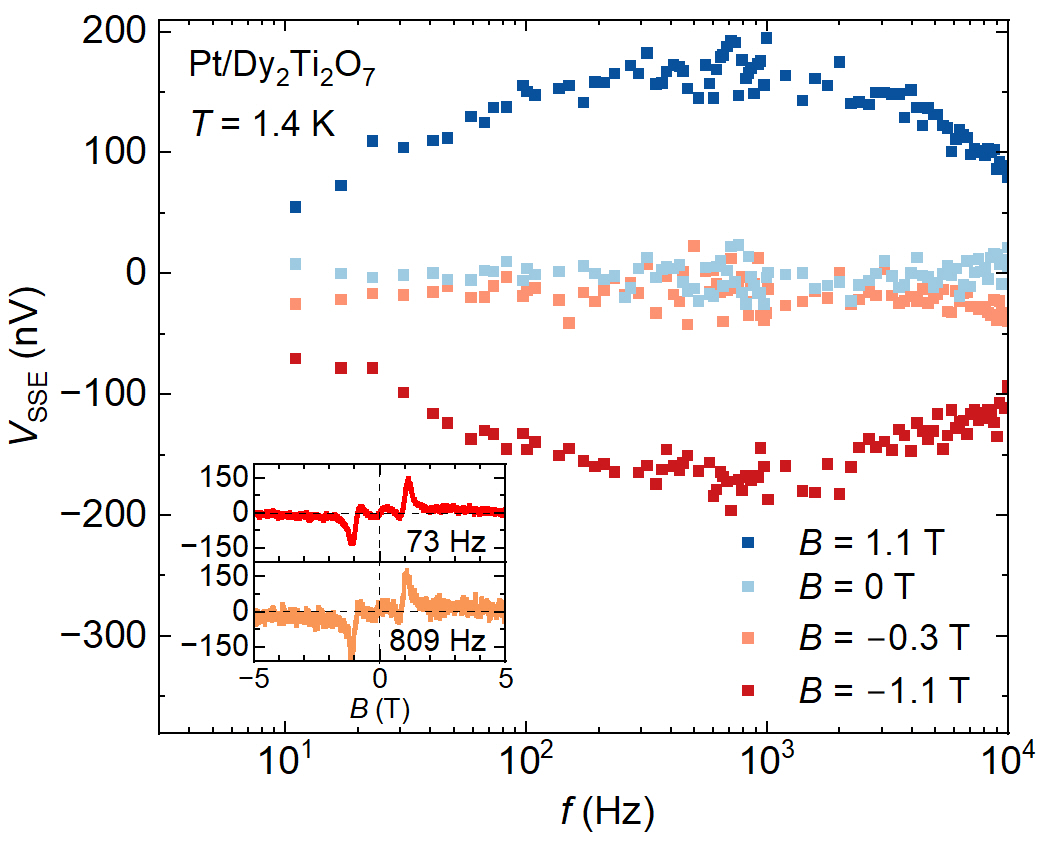}
 \end{center}
\end{figure}
 \normalsize{{\bf Figure 3 $|$ Frequency $f$ dependence of the characteristic peaks in $V_{\rm SSE}$.} $f$ scans of $V_{\rm SSE}$ are shown at selected magnetic fields where characteristic peaks are observed. The $x$-axis is plotted on a logarithmic scale. The inset shows the $B$ dependence of $V_{\rm SSE}$ at selected frequencies for consistency checks.
}
\label{Fig:Fig3}

\newpage
\begin{figure}[ht]
 \begin{center}
 \includegraphics[keepaspectratio, scale=0.7]{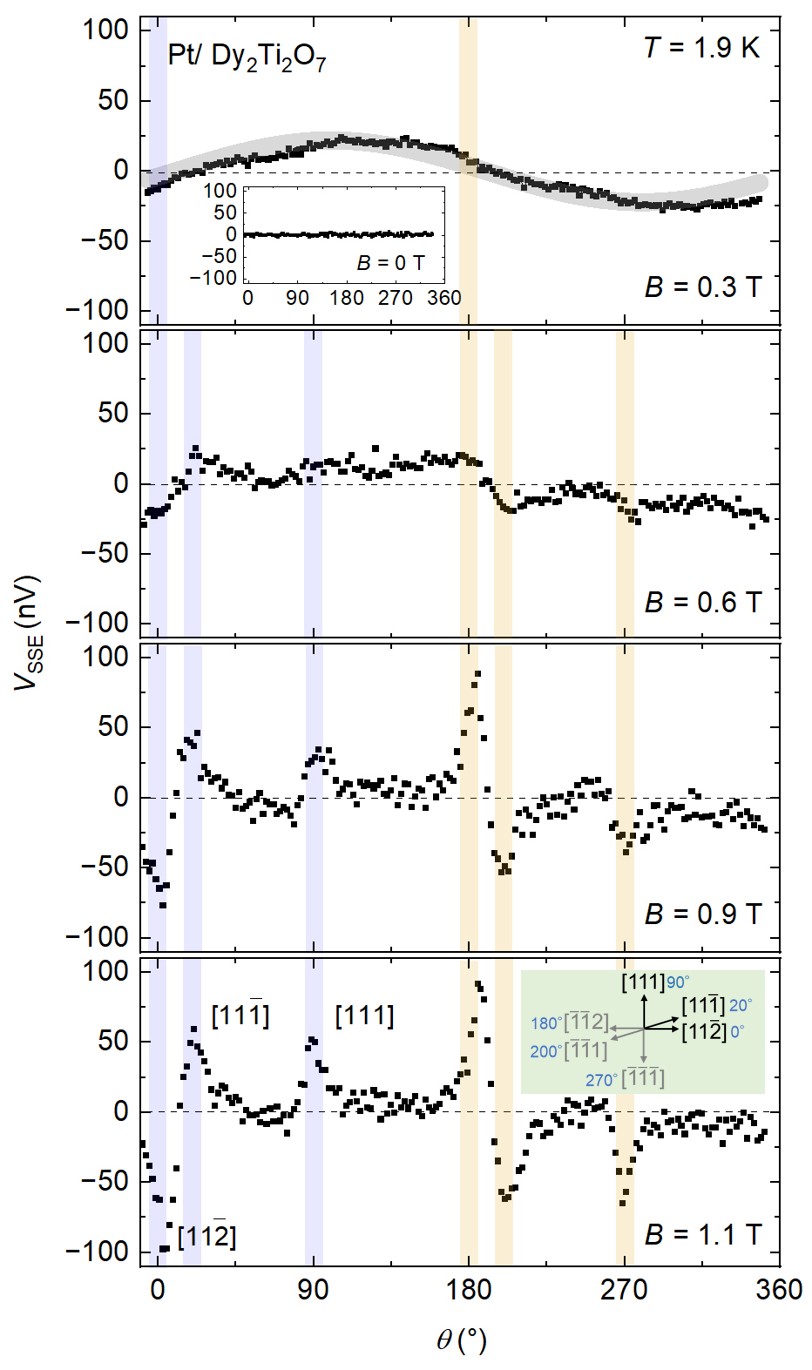}
 \end{center}
\end{figure}
\normalsize{{\bf Figure 4 $|$ In-plane angle ($\theta$) dependence of \VSSE.} The angular dependence of \VSSE{} as a function of $\theta$ within the (1$\bar{1}$0) plane is shown for selected magnetic fields. 
The sample is mounted on a horizontal rotator, which effectively rotates the magnetic field in-plane relative to the sample. 
The inset displays the data at 0 T. 
The angle $\theta$ shown in the figure has already been corrected for an $8^\circ$ offset in the measured rotator angle. This offset was chosen so that the peak associated with $B \parallel [111]$ occurs at $\theta = 90^\circ$.
The offset is likely due to sample misalignment during cutting, polishing, mounting and/or a slight calibration offset of the rotator. 
As noted in Fig.~1, the default SSE configuration corresponds to nominal $\theta = 90^{\circ}$, associated with $B\parallel[111]$.
The thick blue lines indicate $\theta = 0^\circ$, $20^\circ$, and $90^\circ$ with $\pm5^\circ$ error bars, where $B$ is aligned along the high-symmetry axes $[11\bar{2}]$, $[11\bar{1}]$, and $[111]$, respectively. The orange lines denote the corresponding inversion partners, shifted by $180^\circ$. The green panel at the bottom shows a schematic of the orientations of the crystallographic axes. The gray curve represents a $\sin(\theta-\theta_c)$ fit to the \VSSE{} data at 0.3 T, with $\theta_c = 10^\circ$.
The fitted phase shift $\theta_c = 10^\circ$ reflects a possible misalignment of the Hall bar with respect to the crystallographic direction of the \dto{} crystal, as well as the combined uncertainty from the experimental noise and fitting errors. 
We expect the zero crossings of the SSE signal at nominally 0 and 180 $^\circ$ to be shifted by $\theta_c$, which is indeed observed at all fields.
}

\label{Fig:Fig4}

\newpage
\bibliography{DTObib.bib}
\bibliographystyle{naturemag}

\clearpage
\begin{methods}
\section*{Sample preparation} 
A single crystal of \dto{} with the dimensions of 6 mm diameter and 44 mm length was grown in a Crystal Systems Corporation (CSC) optical floating zone furnace. Small rectangular pieces are cut from the single crystal rod. A typical size of the \dto{} substrate used in this study is $4.1 \times 2.3 \times 0.7~mm$. The crystal orientations were checked by a backscattering Laue X-ray diffractometer (Photonic Science) prior to the cutting. The sample is cut using a wire saw (DIDRAS) and subsequently polished with a polishing machine (ECOMET 30, BUEHLER), employing a fine polishing cloth (CHEMOMET, BUEHLER) and a non-crystallizing colloidal silica polishing suspension (MasterMet, BUEHLER). The sample surface roughness is recorded using an Atomic Force Microscopy (Dimension Icon AFM, Bruker). See Extended Data Fig.~5 for a representative AFM picture. The 4-nm-thick Pt (and Ta) Hall bar was fabricated by optical lithography followed by magnetron sputter deposition and lift-off process. The Hall bar consists of a $600 \times 50~\mu\mathrm{m}$ central channel with $20 ~\mu\mathrm{m}$-wide transverse voltage arms. All contact pads are $150 \times 150 ~\mu\mathrm{m}$. See Supplementary Information for more details regarding single crystal growth, surface cleaning and deposition process.

\section*{Control experiment on the nonmagnetic SiO$_{\rm x}$ sample}
A Pt Hall bar of identical size and thickness was sputtered onto commercially purchased (111) Si substrates (0.7 mm thick) with a thermally grown amorphous SiO$_{\rm x}$ surface layer, which exhibits an RMS surface roughness of approximately 700 pm, following the same procedure described above. No additional surface cleaning or surface treatment was applied. Spin Seebeck measurements were performed on the Pt/SiO$_{\rm x}$ sample as a nonmagnetic reference.

\section*{Magnetization measurements}
Magnetization measurements of \dto{} were performed using a superconducting quantum interference device (SQUID) magnetometer in a Magnetic Property Measurement System (MPMS, Quantum Design), equipped with a He$^3$ insert that allows for a base temperature of 0.4 K. The magnetization of the Pt/SiO$_{\rm x}$ device was also measured prior to the spin Seebeck measurements shown in Fig. 2{\bf c}, and no evidence of magnetic contamination was observed.

\section*{SSE measurements}
We measured Spin Seebeck voltages by a standard lock-in technique with a Physical Property Measurement System (PPMS Dynacool, Quantum Design) from 1.4 to 15 K, following the well-established procedure described in Ref.~\citeonline{Vlietstra_2014, Aqeel_2025}. The measurement schematic is shown in Fig.~1{\bf b}. The temperature gradient was generated by AC current heating, via an external Keithley 6221 AC/DC current source. The spin Seebeck voltage \VSSE{} was extracted from the second harmonic component, which exhibits a $-90^\circ$ phase shift relative to the AC current, using lock-in amplifiers (MFLI, Zurich Instruments). The AC current $I_{\rm RMS}$ used in this study ranged from 0.15 to 0.6~mA with a typical frequency of 73.1 Hz. Prior to measurements on \dto{}, a YIG thin film was measured to confirm the sign convention, such that positive \VSSE{} value corresponds to positive magnetic field. To access low temperatures down to 1.4~K, we used the $^3$He option of the PPMS. For angular dependent measurements, we used the horizontal rotator option, which limits the base temperature to 1.9~K. Rotation of the sample with the PPMS horizontal rotator is equivalent to an in-plane rotation of the magnetic field $B$. The angle-dependence data are antisymmetrized, $\bigl[V(+B)-V(-B)\bigr]/2$, to remove a constant offset. Accordingly, three \dto{} devices were used in this study: one Pt/\dto{} device mounted on the $^3$He option, one Pt/\dto{} device mounted on the rotator option, and one Ta/\dto{} device used as a control sample.

\section*{Evaluation of temperature difference based on heat-flow analysis}
For the SSE measurement on the Pt/\dto{} device at the lowest temperature, $T = 1.4~$K, the applied AC current was $I_{\rm RMS} = 0.44$~mA, corresponding to a heating power of $R_{\rm Pt} I_{\rm RMS}^2 = 0.137$~mW, where $R_{\rm Pt}$ is the Pt resistance at the same temperature. The resulting temperature difference between the top surface of the sample and the bottom surface mounted on the sample stage was estimated to be at the order of 0.1~K using a heat-flow analysis based on the model described in Ref.~\citeonline{TKikkawa_2021}. In this analysis, the Pt wire was treated as a heat source embedded at the center of a \dto{} half-cylinder with a radius of 0.7~mm, and the thermal conductivity of \dto{} at $T = 1.4$~K and zero magnetic field was taken from Ref.~\citeonline{Kolland_2013}.

\begin{addendum}
\item[Data Availability]
Source data that support the findings of this study are provided with this paper. 
\end{addendum}
\end{methods}

\clearpage

\newpage
\begin{figure}[ht]
 \begin{center}
  \includegraphics [keepaspectratio,scale=0.4] {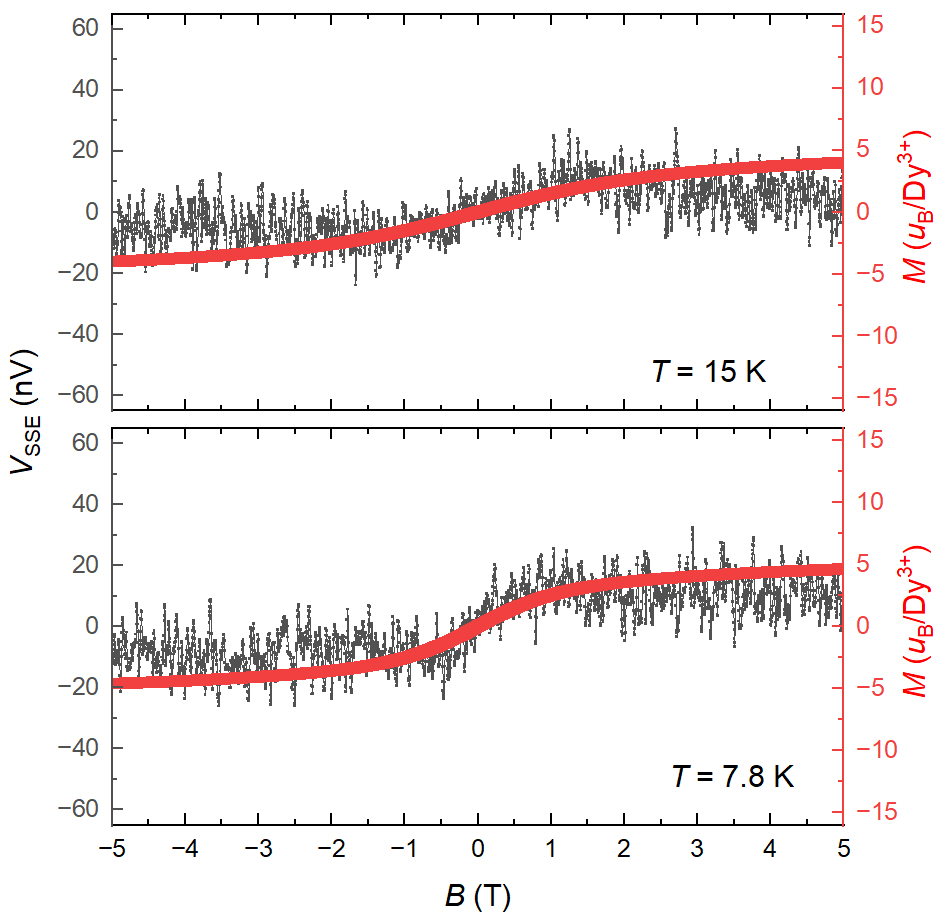}
\end{center}
\end{figure}
 \normalsize{{\bf Extended Data Figure 1 $|$  Comparison of spin Seebeck voltage (\VSSE{}) and magnetization $(M)$ at elevated temperatures.} The left axis shows the \VSSE{} signal, plotted as gray symbols, while the right axis shows the magnetization $M$, plotted as a red curve, measured over the same temperature range.} 
 \label{Fig:FigS4}

\newpage
\begin{figure}[ht]
 \begin{center}
  \includegraphics [keepaspectratio,scale=0.44] {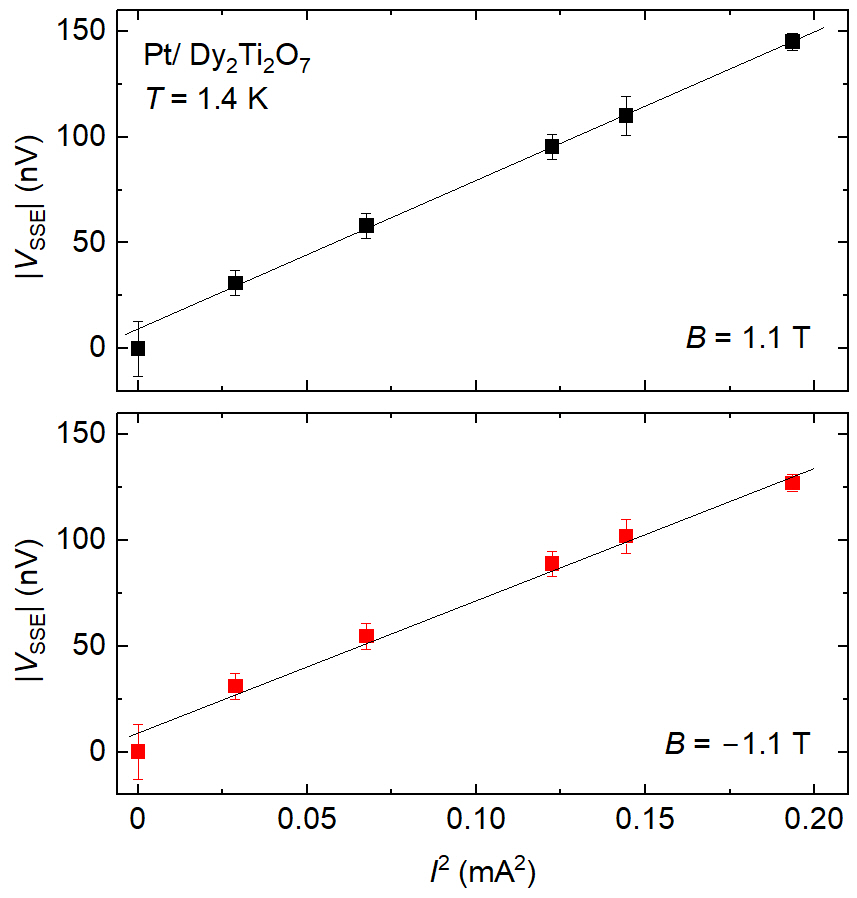}
\end{center}
\end{figure}
 \normalsize{{\bf Extended Data Figure 2 $|$ Quadratic current ($I^2$) dependence of the metamagnetic peak in $|$\VSSE{}$|$.} The metamagnetic peak at both $B = \pm 1.1$ T in $|$\VSSE{}$|$ follows a quadratic dependence on $I$, indicative of its thermal origin. Error bars are visually estimated from the noise fluctuations in the data. Black lines are provided as a guide to the eye.}
 \label{Fig:FigS2}

\newpage
\begin{figure}[ht]
 \begin{center}
  \includegraphics [keepaspectratio,scale=0.27] {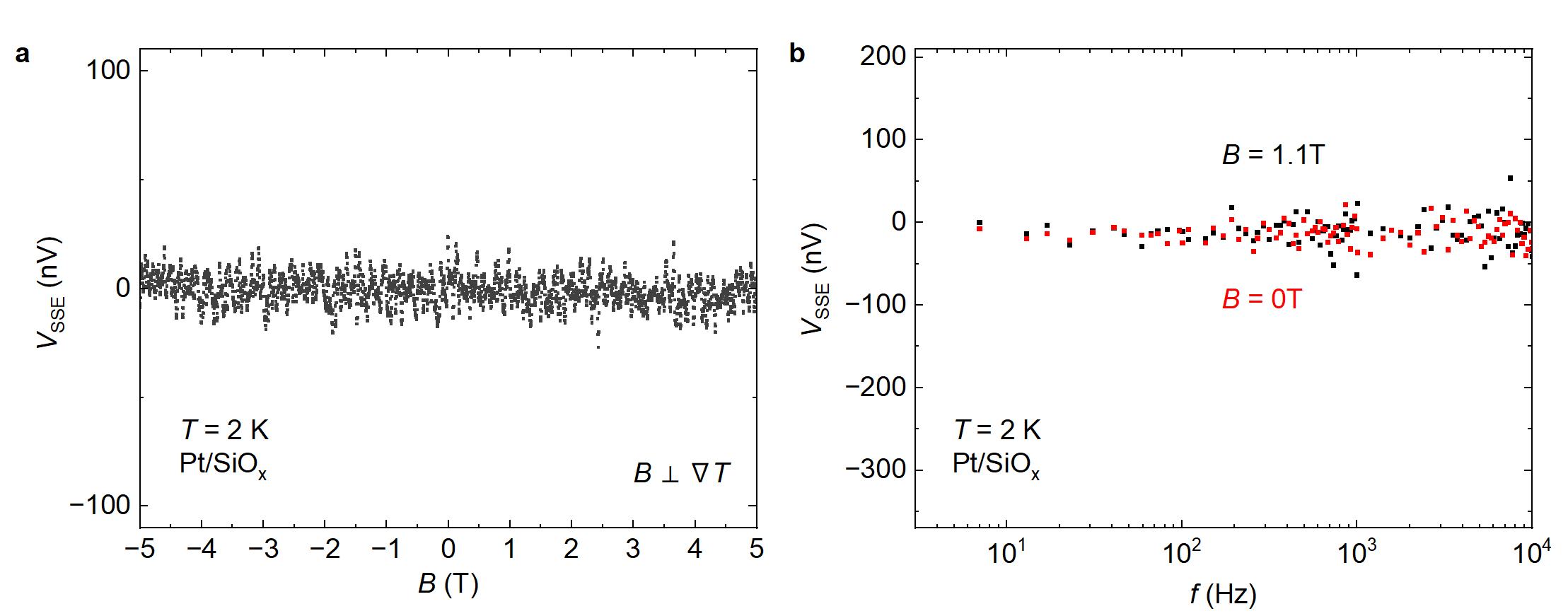}
\end{center}
\end{figure}
\normalsize{{\bf Extended Data Figure 3 $|$ Control experiment for the Pt/SiO$_{\rm x}$ device.} {\bf a}, Magnetic field dependence and {\bf b}, frequency dependence of the \VSSE{} signals in Pt/SiO$_{\rm x}$. The scales are adjusted to match those used in Figs. 2{\bf c} and 3 of the main text.}
 \label{Fig:FigS3}

\newpage
\begin{figure}[ht]
 \begin{center}
  \includegraphics [keepaspectratio,scale=0.26] {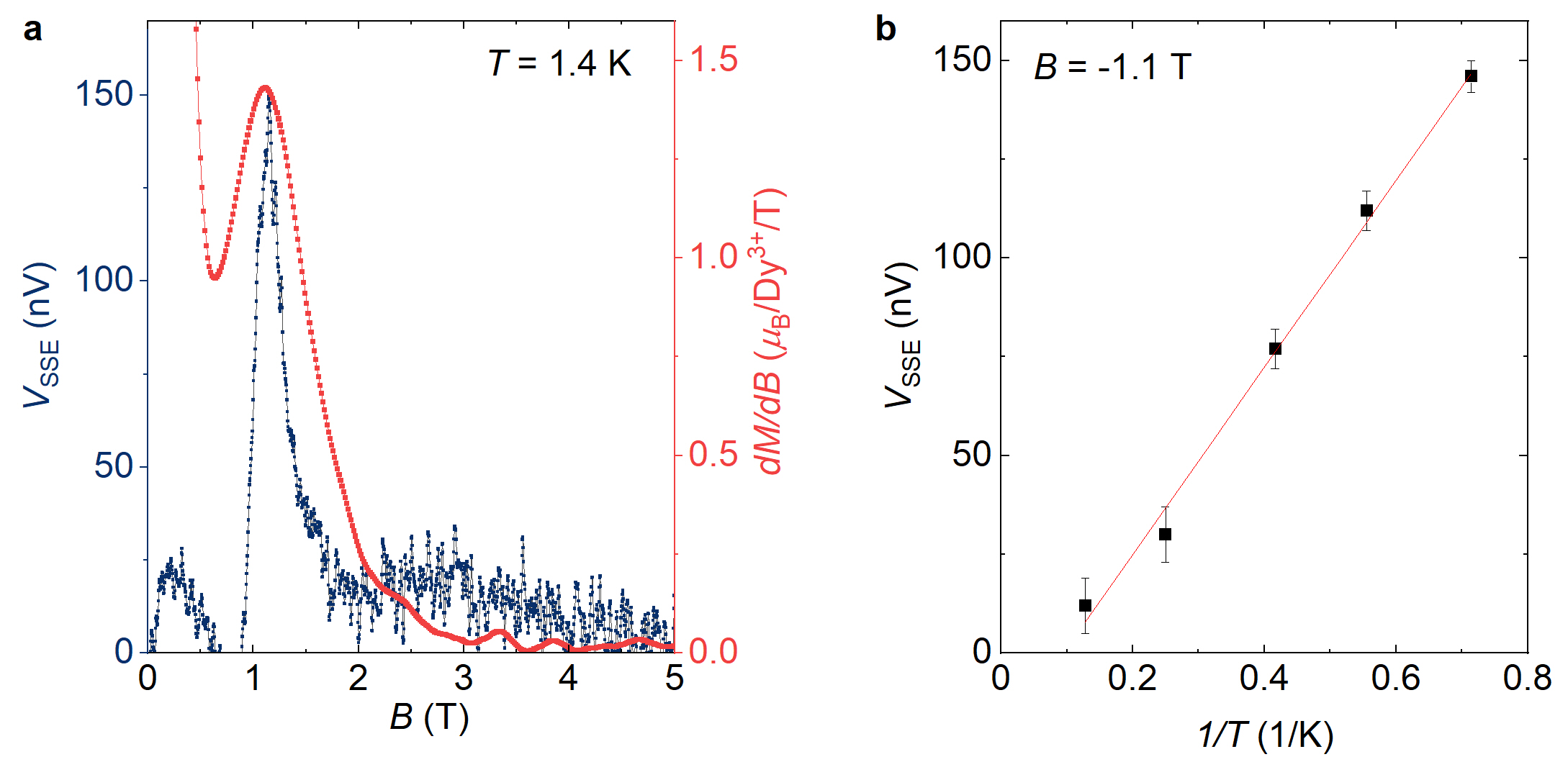}
\end{center}
\end{figure}
 \normalsize{
{\bf Extended Data Figure 4 $|$ Metamagnetic signatures in \VSSE{} and their temperature dependence.} {\bf a}, Comparison of the \VSSE{} peak and the $dM/dB$ peak at 1.4 K under a magnetic field applied along [111]. {\bf b}, Peak height of \VSSE{} at 1.1~T as a function of $1/T$. Error bars are estimated from the noise fluctuations of the measured data. The red line is a linear fit, indicating a $1/T$ dependence of the \VSSE{} peak.
 }
 \label{Fig:FigS6}

\newpage
\begin{figure}[ht]
 \begin{center}
  \includegraphics [keepaspectratio,scale=0.5] {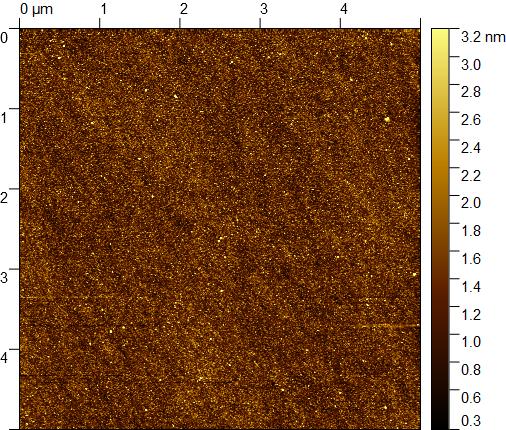}
\end{center}
\end{figure}
 \normalsize{
 {\bf Extended Data Figure 5 $|$ AFM image of \dto{} surface before lithography and deposition.} 
 The root-mean-square (RMS) surface roughness $S_{\rm q}$ is $\sim$ \(430~\mathrm{pm}\).
 }
 \label{Fig:FigS1}

\end{document}